\documentclass[11pt,epsf]{article}
\usepackage{amsmath}
\usepackage{amsfonts}
\usepackage{amssymb}
\usepackage{graphicx}
\usepackage{color}
\usepackage{comment}

\newtheorem{theorem}{Theorem}

\newcommand{\hU}{{\hat{U}}}
\newcommand{\hX}{{\hat{X}}}
\newcommand{\tX}{{\tilde{X}}}
\newcommand{\hY}{{\hat{Y}}}
\newcommand{\hV}{{\hat{V}}}
\newcommand{\hW}{{\hat{W}}}
\newcommand{\hZ}{{\hat{Z}}}

\newcommand {\dfn} {\stackrel{\Delta} {=}}

\newcommand {\reals} {{\rm I\!R}}

\newcommand {\bu} {\mbox{\boldmath $u$}}
\newcommand {\bv} {\mbox{\boldmath $v$}}
\newcommand {\bw} {\mbox{\boldmath $w$}}
\newcommand {\bx} {\mbox{\boldmath $x$}}
\newcommand {\by} {\mbox{\boldmath $y$}}

\newcommand {\bE} {\mbox{\boldmath $E$}}

\newcommand{\calA}{{\cal A}}

\newcommand{\calE}{{\cal E}}

\newcommand{\calP}{{\cal P}}

\newcommand{\calS}{{\cal S}}
\newcommand{\calT}{{\cal T}}
\newcommand{\calU}{{\cal U}}
\newcommand{\calV}{{\cal V}}
\newcommand{\calW}{{\cal W}}
\newcommand{\calX}{{\cal X}}
\newcommand{\calY}{{\cal Y}}
\newcommand{\calZ}{{\cal Z}}
\allowdisplaybreaks

\begin{document}
\thispagestyle{empty}
\title{Finite--State Source--Channel Coding for Individual Source Sequences
with Source Side Information at the Decoder}
\author{Neri Merhav
}
\date{}
\maketitle

\begin{center}
The Viterbi Faculty of Electrical and Computer Engineering\\
Technion - Israel Institute of Technology \\
Technion City, Haifa 32000, ISRAEL \\
E--mail: {\tt merhav@ee.technion.ac.il}\\
\end{center}
\vspace{1.5\baselineskip}
\setlength{\baselineskip}{1.5\baselineskip}

\begin{abstract}
We study the following semi--deterministic setting of the joint source--channel
coding problem: a
deterministic source sequence (a.k.a.\ individual sequence) is transmitted via
a memoryless channel, using delay-limited encoder and decoder, which are both
implementable by periodically--varying finite-state machines, and the decoder is granted 
with access to side
information, which is a noisy
version of the source sequence. We first derive a lower bound on the
achievable expected distortion in terms of the empirical statistics of the source
sequence, the number of states of the encoder, the number of states of the
decoder, their period, and the overall delay. The bound is shown to be asymptotically
achievable by universal block codes in the limit of long blocks. 
We also derive a lower bound to the best achievable excess--distortion probability and discuss
situations where it is achievable. Here, of course, source coding and channel coding
cannot be completely separated without loss of optimality. Finally, we outline a few
extensions of the model considered, such as: (i) incorporating a common reconstruction constraint,
(ii) availability of side information at both ends, and (iii) extension to the Shannon channel with causal state information at the encoder.
This work both extends and improves on earlier work of
the same flavor (Ziv 1980, Merhav 2014), which focused only on the expected
distortion, without side information at either end, and without the above mentioned
additional ingredients.\\

\end{abstract}

\clearpage
\section{Introduction}

In a collection of works that appeared during the late
seventies and eighties of the previous century, Ziv
\cite{Ziv78}, \cite{Ziv80}, \cite{Ziv84}, and Ziv and Lempel
\cite{LZ86}, \cite{ZL78},
have established a fascinating theory of universal source coding for
deterministic sequences (a.k.a.\ individual sequences) by means of
encoders/decoders that are implementable using finite--state machines.
Specifically, in \cite{Ziv78} Ziv
addressed the issue of fixed--rate, universal (nearly)
lossless compression of deterministic source sequences
using finite--state encoders and decoders,
which was later further developed to the
celebrated Lempel--Ziv algorithm \cite{LZ86}, \cite{ZL78}.
In \cite{Ziv80}, the model setting of \cite{Ziv78} was broadened to
lossy transmission over both clean and noisy channels (subsections II.A and
II.B therein, respectively), where in the noisy case, the channel was modeled as
an ordinary, probabilistic memoryless channel, as opposed to the source sequence,
that was still assumed deterministic. Henceforth, we will refer to this type of setting
as a {\em semi--deterministic} setting, similarly as in \cite{me14}.

Subsequently, the results of the first part of \cite{Ziv80} (clean channels) were
further elaborated in other directions, such as exploiting side
information in a scenario of almost lossless source coding, where the side
information is modeled too as being deterministic \cite{Ziv84}, i.e., a
deterministic analogue of Slepian--Wolf coding 
was investigated in \cite{Ziv84}. More than two decades later, this 
setup was extended to the lossy case
\cite{MZ06}, that is, a semi--deterministic counterpart of Wyner--Ziv
coding, where the source to be compressed is deterministic, but the 
side information available to the decoder is generated from the
source sequence via a discrete memoryless channel (DMC). 
The model of \cite{MZ06} was still a pure source--coding model,
where the main channel was clean, and the encoding model allowed
variable--length coding.
In \cite{me14}, a few inaccuracies in the coding theorem for noisy
channels in \cite[Subsection II.B]{Ziv80} were corrected, and it was
strengthened and refined from several aspects. Among other things, in
\cite{me14}, only the decoder was assumed to be a finite--state machine, while the
encoder was allowed to be rather general (as opposed to \cite{Ziv80}, where the
encoder was assumed to be a finite--state machine too). Also, the finite--state decoder
of \cite{me14} was allowed to be periodically time--varying 
with a given period length, $\ell$, along with a modulo--$\ell$ time counter (clock).

In this work, we further develop the findings of \cite{me14} and \cite{MZ06} in a few directions, and at the
same time, we also take the opportunity to correct 
some (minor) imprecisions in \cite{me14} and improve the rigor of the derivation, as well as the tightness of the converse bound.
In our present model, we are back to assume that both the encoder and the decoder are
finite--state machines (similarly as in \cite{Ziv80} and \cite{Ziv84}), but as in \cite{me14}, we
allow them to be periodically 
time--varying (with the same period), having a limited delay, and we 
also allow the decoder to access
side information, which is 
a noisy version (corrupted by a DMC) of the deterministic source sequence to be
conveyed -- see Fig.\ \ref{fig1}. In other words, it is a semi--deterministic setting 
of a joint source--channel coding problem that
combines the semi--deterministic version of the Wyner--Ziv (W-Z) source model with the
DMC, in analogy to the purely stochastic version of this 
model \cite{SVZ98}. The W-Z channel can be motivated by the uncoded transmission of the
systematic part of a systematic code (see also \cite{MS03}, \cite{SVZ98}). 

At first glance, one might wonder about this asymmetric modeling approach of the semi-deterministic setting
(both here and in the earlier works, \cite{me14}, \cite{MZ06}, \cite{Ziv80}), where 
the source is regarded completely deterministically, without any statistical
assumptions, while the channels
(namely, both the main channel and the Wyner--Ziv channel)
are modeled probabilistically, exactly as in the classical tradition of the information theory literature.
The motivation for this distinction, is that in many frequently encountered situations, 
the channels obey some relatively well--understood physical 
laws that govern the underlying noise processes, which can be reasonably well be modeled probabilistically,
whereas the
source to be conveyed is very different in nature. 
Indeed, in many applications, the source is a 
man--made data file (or a group of files), generated using artificial means. These include
computer--generated images and video streams, texts of various types, 
audio signals (such as music), sequences of output results from computer calculations, 
and any combinations of those. It is simply inconceivable to use ordinary probabilistic models for such sources.

For the above described semi-deterministic model, we first derive a lower bound to the minimum achievable expected distortion 
between the source and its reconstruction at the decoder. 
Our main result is a lower bound to the best
achievable distortion, which depends on: (i) the given source sequence, (ii) the capacity of the
main channel, (iii), the W-Z channel, (iv) the period $\ell$, (v) the number of states of the encoder,
$s_{\mbox{\tiny e}}$, and (vi) the number of states of the decoder, $s_{\mbox{\tiny d}}$. It turns out that the lower bound
depends on $s_{\mbox{\tiny e}}$ {\em very differently} than on $s_{\mbox{\tiny d}}$: The dependence on $s_{\mbox{\tiny e}}$
is much weaker and it completely vanishes as the length $n$ of the source sequence tends to infinity. In other words, as far as the
lower bound is concerned, $s_{\mbox{\tiny d}}$ is a much more significant resource than $s_{\mbox{\tiny e}}$. 
This asymmetric
behavior of the lower bound is interesting and not trivial.
The bound is also tighter than in \cite{Ziv80}. 
It can be
universally asymptotically achieved by separate W-Z source coding
and channel coding, using long block codes, provided that both $d$
and $\log s_{\mbox{\tiny d}}$ are small compared to $\ell$.
 
In addition to the the expected distortion, we also address a related, but different figure of merit --
the probability of excess distortion,
similarly as in \cite{Csiszar82} and \cite{Marton74}. We first derive a lower
bound to this probability for the simpler model setting without side information,
as in \cite{me14} and \cite{Ziv80}. We relate it to the probability of excess
distortion in the purely probabilistic
setting \cite{Csiszar80}, \cite{Csiszar82}, and then 
discuss when this bound is asymptotically
achievable. Subsequently,
we extend the scope to the above--described model with decoder side information.
Achievability is discussed too, but it should be pointed out that
our emphasis, in this work, is on fundamental limits and lower bounds, more than on 
achievability. 

In the last part of this work, we discuss a few variants of 
the model, where more explicit results can be stated, such as
the case where the source side information 
is available at the encoder too. The case where a common reconstruction
constraint is imposed, following \cite{Steinberg09}, is also presented. Finally, we discuss an extension
where the ordinary DMC is replaced by the Shannon channel with causal state information at the encoder \cite{Shannon58}.

The outline of the remaining part of this article is as follows.
In Section \ref{ncpf}, we establish notation conventions and formalize the problem setting and the objectives.
In Section \ref{pre}, we provide a few additional definitions in order to establish the
preparatory background needed to state the main results, and we also provide a preliminary result, for the case
of an ordinary DMC and without side information. In Section \ref{wz}, we provide the extension that incorporates
source side information, and finally, in Section \ref{ext}, we outline a few modifications and extensions
of our setting as described in the previous paragraph.

\section{Notation Conventions and Problem Formulation}
\label{ncpf}

\subsection{Notation Conventions}

Throughout the paper, random variables will be denoted by capital
letters, specific values they may take will be denoted by the
corresponding lower case letters, and their alphabets
will be denoted by calligraphic letters. Similarly, random
vectors, their realizations, and their alphabets, will be denoted,
respectively, by capital letters, the corresponding lower case letters,
and calligraphic letters, all superscripted by their dimensions. For
example, the random vector $Y^n=(Y_1,\ldots,Y_n)$, ($n$ -- positive
integer) may take a specific vector value $y^n=(y_1,\ldots,y_n)$
in $\calY^n$, the $n$--th order Cartesian power of $\calY$, which is
the alphabet of each component of this vector. An infinite sequence will be denoted by the bold face font,
for example, $\bu=(u_1,u_2,\ldots)$. The notation $\bu_i$, on the other hand, will be used to denote the $i$--th $\ell$--block
$(u_{i\ell+1},u_{i\ell+2},\ldots,u_{i\ell+\ell})$. For $i\le j$,
($i$, $j$ -- positive integers), $x_i^j$ will denote the segment
$(x_i,\ldots,x_j)$, where for $i=1$ the subscript will be omitted.
If, in addition, $j=1$, the superscript will be omitted too, and the notation will be simply $x$.

Owing to the semi-deterministic modeling approach, 
we distinguish between two kinds of random variables: ordinary random variables (or vectors), governed by certain given probability
distributions (like the channel output vector), 
and auxiliary random variables that emerge from empirical distributions associated with certain sequences. Random
variables of the second kind will be denoted using `hats'. For example, consider a deterministic sequence $u^n=(u_1,\ldots,u_n)$.
Then, $\hU^\ell=(\hU_1,\ldots,\hU_\ell)$ designates an auxiliary random vector, `governed' by the empirical distribution extracted
from the non-overlapping $\ell$--blocks of $u^n$, provided that $\ell$ divides $n$ (this empirical distribution will 
be defined precisely in the sequel).
We denote this empirical distribution by
$P_{\hU^\ell}=\{P_{\hU^\ell}(u^\ell),~u^\ell\in\calU^\ell\}$. 
The use of empirical distributions, however, will not be limited to
deterministic sequences only. It could be defined also for realizations of 
random sequences. For example, if $Y^n$ is a random sequence,
$\hY^\ell$ would designate the auxiliary random $\ell$-vector associated with a given realization, $y^n$, of $Y^n$. In this case, the
empirical distribution, $P_{\hY^\ell}$, is of course, itself random, but it may converge to the true $\ell$--th order
distribution of $Y^\ell$, $P_{Y^\ell}$, under certain conditions. Information measures, like entropies, conditional entropies, 
divergences, and mutual informations, will be denoted according to the conventional rules of the information theory literature,
where it should be kept in mind that these measures may involve both ordinary and auxiliary random variables. For example,
$I(\hU^\ell;Y^\ell)$ and $H(Y^\ell|\hU^\ell)$ are, respectively, the mutual information 
and the conditional entropy induced by the empirical distribution of $\hU^\ell$, $P_{\hU^\ell}$, and
the conditional distribution, $P_{Y^\ell|\hU^\ell}$, 
of the ordinary random vector, $Y^\ell$, given the auxiliary random vector, $\hU^\ell$. The conditional divergence,
$D(Q_{Y|\hX}\|P_{Y|\hX}|P_{\hX})$, will be understood to be given by
\begin{equation}
	D(Q_{Y|\hX}\|P_{Y|\hX}|P_{\hX})=\sum_{x\in\calX}P_{\hX}(x)\sum_{y\in\calY}
	Q_{Y|\hX}(y|x)\log\frac{Q_{Y|\hX}(y|x)}{P_{Y|\hX}(y|x)},
\end{equation}
where logarithms, here and throughout the sequel, will be understood to be taken to the base 2, unless specified otherwise.

\begin{figure}[h]
\hspace*{1cm}\input{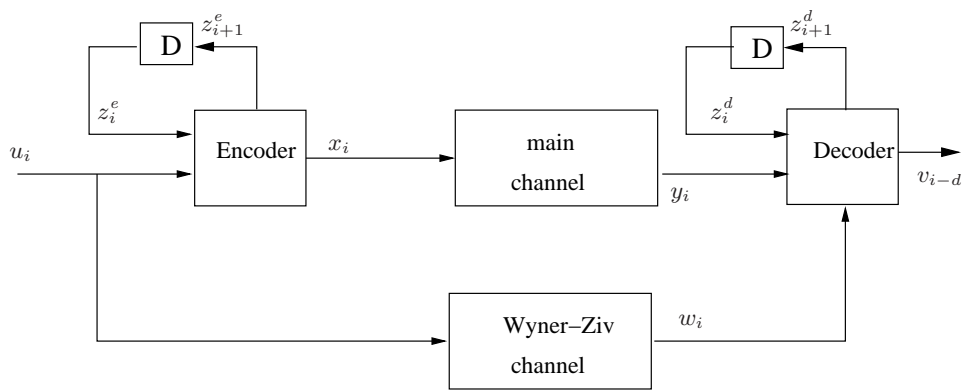}
\caption{Source--channel 
with side information at the decoder according
to the formal model description in Subsection 2.2. Both the encoder and the decoder are
finite--state machines with an overall delay, $d$.
The label ``D'', in each of the feedback loops of the encoder and the decoder,
designates a device that introduces a delay of one time unit, thus passing from the
next state (at time $i+1$) to the current state (at time $i$).}
\label{fig1}
\end{figure}

\subsection{Problem Formulation}

Referring to Fig.\ \ref{fig1},
let $\bu=(u_1,u_2,\ldots)$ be a deterministic source sequence of symbols in a finite alphabet
$\calU$ of cardinality $|\calU|=\alpha$. 
The sequence $\bu$ 
is encoded using a periodically time--varying finite--state
encoder, whose output is $\bx=(x_1,x_2,\ldots)$, where $x_i\in\calX$, $\calX$
being another
finite alphabet,
of size $|\calX|=\beta$. More precisely, the encoder obeys the following
equations
\begin{eqnarray}
	\label{fsme}
	t&=&i~\mbox{mod}~\ell\\
x_i&=&f_t(u_i,z_i^{\mbox{\tiny e}}),\\
z_{i+1}^{\mbox{\tiny e}}&=&g_t(u_i,z_i^{\mbox{\tiny e}}),
\end{eqnarray}
where $i=1,2,\ldots$, $z_i^{\mbox{\tiny e}}$ is the encoder state at time $i$, which takes
values in a finite set, $\calZ^{\mbox{\tiny e}}$, of size $s_{\mbox{\tiny e}}$.
The functions $f_t$ and $g_t$ are, respectively, the periodically
time--varying {\em output function} and
the {\em next--state
function} of the encoder. The length of the period is $\ell$.
The sequence
$\bx$ is fed into a DMC, characterized by the single-letter
transition probabilities $\{P_{Y|X}(y|x),~x\in\calX,~y\in\calY\}$, where $\calY$ is
a finite alphabet of size $|\calY|=\gamma$. 
The channel output
$\by=(y_1,y_2,\ldots)$ is fed into a periodically time--varying finite--state decoder, which
is defined by
\begin{eqnarray}
	\label{fsmd}
	t&=&i~\mbox{mod}~\ell,~~~~~~~~~~i=1,2,\ldots\\
v_{i-d}&=&f_t'(w_i,y_i,z_i^{\mbox{\tiny d}}), ~~~~~~~~i=d+1,d+2,\ldots\\
z_{i+1}^{\mbox{\tiny d}}&=&g_t'(w_i,y_i,z_i^{\mbox{\tiny d}}), ~~~~~~~~i=1,2,\ldots
\end{eqnarray}
where $z_i^{\mbox{\tiny d}}\in\calZ^{\mbox{\tiny d}}$ is the decoder state at
time $i$, $\calZ^{\mbox{\tiny d}}$ being a finite
set of states of size $s_{\mbox{\tiny d}}$, $w_i\in\calW$ is the side information at time $i$, and
$v_{i-d}$ is the reconstructed version of the source sequence, delayed by $d$
time units ($d$ -- positive integer). The reconstruction alphabet is $\calV$
of size $\delta$. The side information sequence, $\bw=(w_1,w_2,\ldots)$, is generated from $\bu$
by means of a DMC, characterized by a matrix of transition probabilities, $P_{W|U}=\{P_{W|U}(w|u),~u\in\calU,~w\in\calW\}$.
The functions $f_t'$ and $g_t'$ are, respectively, the output function and the
next--state function of the decoder.

Let 
$V^n=(V_1,\ldots,V_n)$,
$W^n=(W_1,\ldots,W_n)$,
and $Y^n=(Y_1,\ldots,Y_n)$,
designate random vectors pertaining to the variables
$v^n=(v_1,\ldots,v_n)$,
$w^n=(w_1,\ldots,w_n)$,
and $y^n=(y_1,\ldots,y_n)$,
respectively, where the randomness stems from 
the main channel and the W--Z channel. The vector $u^n$ clearly does not have a stochastic
counterpart. The same comment applies also 
to $x^n$ since the encoder is assumed deterministic.

The objectives of the paper are the following: given the source sequence $u^n$, the channel,
$P_{Y|X}$, and the W--Z channel $P_{W|U}$, the numbers of encoder and decoder states, 
$s_{\mbox{\tiny e}}$ and $s_{\mbox{\tiny d}}$, the period, $\ell$, and the allowed delay, $d$,
and given a single--letter distortion function $\rho:\calU\times\calV\to\reals^+$, we wish to find non--trivial lower bounds to:
\begin{enumerate}
	\item The expected distortion, $\frac{1}{n}\sum_{i=1}^n\bE\{\rho(u_i,V_i)\}$, and
	\item The probability of excess distortion, $\mbox{Pr}\{\sum_{i=1}^n \rho(u_i,V_i)\ge nD\}$, where $D> 0$ 
		is a constant larger than the
		best achievable normalized expected distortion.
\end{enumerate}
In some instances of the problem, we will also discuss asymptotic achievability.

\section{Background and Preliminary Results}
\label{pre}

Before moving forward to our main results, we need a few more definitions, as well as some background on the
relevant results from \cite{me14}. 
We conclude this section with a preliminary result on the excess distortion probability
in the case of an ordinary DMC and without source--related side information at the decoder.

Let $\ell$ divide $n$ and consider the segmentation of all relevant sequences into
$n/\ell$ non--overlapping blocks of length $\ell$, that is,
\begin{equation}
u^n=(\bu_0,\bu_1,\ldots,\bu_{n/\ell-1}),
~~~~~~~~\bu_i=(u_{i\ell+1},u_{i\ell+2},\ldots,u_{i\ell+\ell}),~~~~i=0,1,\ldots,n/\ell-1,
\end{equation}
and similar definitions for $v^n$, $w^n$, $x^n$, and $y^n$, where
$v_{n-d+1}, v_{n-d+2},\ldots,v_n$ (which are not yet
reconstructed at time $t=n$) are defined as 
arbitrary symbols in $\calV$.
Let us define the empirical joint probability mass function
\begin{eqnarray}
& &P_{\hU^\ell\hV^\ell\hW^\ell\hX^\ell\hY^\ell\hZ^{\mbox{\tiny
e}}\hZ^{\mbox{\tiny d}}}(u^\ell,v^\ell,w^\ell,x^\ell,y^\ell,z^{\mbox{\tiny
e}},z^{\mbox{\tiny d}})\nonumber\\
&=&\frac{\ell}{n}\sum_{i=0}^{n/\ell-1}
	1\{\bu_i=u^\ell,\bv_i=v^\ell,\bw_i=w^\ell,\bx_i=x^\ell,\by_i=y^\ell,z_{i\ell+1}^{\mbox{\tiny
	e}}=z^{\mbox{\tiny e}},z_{i\ell+1}^{\mbox{\tiny d}}=z^{\mbox{\tiny d}}\},
\end{eqnarray}
where $1\{\ldots\}$ is the indicator function of the combination of events indicated in its argument.
Clearly, since $P_{\hU^\ell\hV^\ell\hW^\ell\hX^\ell\hY^\ell\hZ^{\mbox{\tiny
e}}\hZ^{\mbox{\tiny d}}}$ is a legitimate probability
distribution, all the rules of manipulating information measures (the chain
rule, conditioning reduces entropy, etc.) hold as usual. Marginal and conditional marginal 
distributions associated with subsets of the set of random variables,
$(\hU^\ell,\hV^\ell,\hW^\ell,\hX^\ell,\hY^\ell,\hZ^{\mbox{\tiny
e}},\hZ^{\mbox{\tiny d}})$, which
are derived from $P_{\hU^\ell\hV^\ell\hW^\ell\hX^\ell\hY^\ell\hZ^{\mbox{\tiny
e}}\hZ^{\mbox{\tiny d}}}$, will be denoted using the conventional notation, for example,
$P_{\hU^\ell\hY^\ell}$ is the joint empirical distribution of $(\hU^\ell,\hY^\ell)$,
$P_{\hY^\ell|\hX^\ell\hZ^{\mbox{\tiny e}}}$ is the conditional empirical
distribution of $\hY^\ell$ given $(\hX^\ell,\hZ^{\mbox{\tiny e}})$, and so on.

We now define the W-Z rate--distortion function \cite{WZ76} 
of the source $P_{\hU^\ell}$ with respect to the {\em real} side--information
channel $P_{W^\ell|\hU^\ell}$ (as opposed to the empirical 
side--information channel, $P_{\hW^\ell|\hU^\ell}$) according to
\begin{equation}
	R_{\hU^\ell|W^\ell}^{\mbox{\tiny WZ}}(D)=\frac{1}{\ell}\min\{I(\hU^\ell;A)-I(W^\ell;A)\}\equiv\min 
	\frac{1}{\ell}I(\hU^\ell;A|W^\ell),
\end{equation}
where both minima are taken over all conditional distributions, $\{P_{A|\hU^\ell}\}$, 
such that $A\to\hU^\ell\to W^\ell$ is a Markov chain
and $\min_{\{G:\calA\times\calW^\ell\to\calV^\ell\}}\bE\{\rho(\hU^\ell,G(A,W^\ell))\}\le \ell\cdot D$, where
$\rho(u^\ell,v^\ell)$ is defined additively as $\sum_{i=1}^\ell\rho(u_i,v_i)$ and $A$ is an auxiliary RV whose alphabet size is
$|\calA|=\alpha^\ell+1$. It follows from these definitions that if $W^\ell\to\hU^\ell\to A$ is a Markov chain,
then
\begin{equation}
	I(\hU^\ell;A)-I(W^\ell;A)\equiv I(\hU^\ell;A|W^\ell)\ge\ell\cdot
	R_{\hU^\ell|W^\ell}[\Delta(\hU^\ell|W^\ell,A)/\ell],
\end{equation}
where
\begin{equation}
	\Delta(\hU^\ell|W^\ell,A)\dfn\min_{G:\calW^\ell\times\calA\to\calV^\ell}\bE\{\rho(\hU^\ell,
	G(W^\ell,A))\},
\end{equation}
where $\calA$ is the alphabet of $A$ and
the expectation is taken with respect to (w.r.t.) \ $P_{\hU^\ell W^\ell A}=P_{\hU^\ell A}\times P_{W^\ell|\hU^\ell}$.
Clearly, if $W^\ell$ is independent of $\hU^\ell$, that is, $P_{W^\ell|\hU^\ell}(\cdot|u^\ell)$ is the
same for all $u^\ell\in\calU^\ell$, then $R_{\hU^\ell|W^\ell}^{\mbox{\tiny WZ}}(D)$ degenerates to the ordinary rate--distortion
function, which will be denoted by $R_{\hU^\ell}(D)$. In the sequel, we will also refer to the conditional rate--distortion,
of $\hU^\ell$ given $W^\ell$, which will be denoted by $R_{\hU^\ell|W^\ell}(D)$, 
where $W^\ell$ is available to both encoder and decoder. This function
is also given by the minimum of $I(\hU^\ell;A|W^\ell)/\ell$, except that the Markov condition is dropped \cite{Zamir96}.
The corresponding distortion--rate functions, $D_{\hU^\ell|W^\ell}^{\mbox{\tiny WZ}}(R)$,
$D_{\hU^\ell}(R)$, and
$D_{\hU^\ell|W^\ell}(R)$, are the inverse functions of
$R_{\hU^\ell|W^\ell}^{\mbox{\tiny WZ}}(D)$,
$R_{\hU^\ell}(D)$, and
$R_{\hU^\ell|W^\ell}(D)$, respectively.

For the given channel, $P_{Y|X}$, 
we denote by $C_{P_{Y|X}}(\Gamma)$ the channel capacity with a transmission cost constraint, $\sum_{i=1}^n\bE\{\phi(X_i)\}\le n\Gamma$ 
($\phi(\cdot)$ being the single--letter transmission cost function), that is
\begin{equation}
	C_{P_{Y|X}}(\Gamma)=\max_{P_X:~\bE\{\phi(X)\}\le\Gamma} I(X;Y).
\end{equation}
When the channel, $P_{Y|X}$, is clear from the context, the
subscript ``$P_{Y|X}$'' will be omitted, and the notation will be simplified to $C(\Gamma)$.

In \cite{me14}, we considered the simpler case
without decoder side information, $\bw$, related to the source, and where only the decoder is limited
to $s$ states.
One of the main results of \cite{me14} (in particular,
Theorem 1 therein) is a lower bound to the expected distortion, which has the following form:
\begin{equation}
\label{mainsemi}
	\frac{1}{n}\sum_{i=1}^n\bE\{\rho(u_i,V_i)\}\ge D_{\hU^\ell}\left(C(\Gamma)+\zeta(s,d,\ell)+
	\epsilon(\ell,n)\right),
\end{equation}
where
\begin{equation}
	\epsilon(\ell,n)\dfn\frac{(\alpha\beta)^\ell \log \gamma}{\sqrt{n}}+o\left(\frac{1}{\sqrt{n}}\right),
\end{equation}
and $\zeta(s,d,\ell)$ is a certain function\footnote{The exact form of this function is immaterial for the purpose of this
discussion. In fact, the formula of $\zeta(s,d,\ell)$, 
given in \cite{me14} is somewhat imprecise, and so, we both correct and extend it in this work. Nevertheless,
the property $\lim_{\ell\to\infty}\zeta(s,d,\ell)=0$ remains valid.}
with the property $\lim_{\ell\to\infty}\zeta(s,d,\ell)=0$.
As discussed in \cite{me14}, it is interesting that 
the distortion bound depends on $u^n$ only via its $\ell$--th order
empirical distribution, $P_{\hU^\ell}$, where, as defined above, $\ell$ is
length of the period. It is
also discussed in that work that the term
$\zeta(s,d,\ell)$, on the right--hand side, can be thought of as an ``extra capacity'' term, that
is induced by 
the memory of the encoding--decoding system (encapsulated in the
state) and the allowed delay, but its effect is diminished
when $\ell$ is chosen large.
The bound can then be asymptotically approached by separate source-- and channel coding, 
using long block codes (see also the achievability scheme described in detail in the discussion of Section 4 below). 
On the other hand, by letting $\ell$ grow,
one also affects the distortion--rate function, 
$D_{\hU^\ell}(\cdot)$, and so, the overall effect of $\ell$ is not trivial to assess in
general. 

We now state our preliminary result on the excess distortion probability for the case of DMC and without any side information.
\begin{theorem}
	\label{thm1}
	Assume that $\rho_{\max}=\max_{u,v}\rho(u,v) < \infty$. Then, under
the assumptions of \cite{me14},
for a given $u^n$, an arbitrary encoder and a finite--state decoder with $s$ states and overall delay $d$,
\begin{eqnarray}
	\mbox{Pr}\left\{\sum_{i=1}^n\rho(u_i,V_i)\ge nD\right\}&\ge&\sup_{\Delta>0}
        \left[\frac{\Delta}{\rho_{\max}-D}-o(n)\right]\times\nonumber\\
	& &\exp\left\{-(n+d)E_{\mbox{\tiny sp}}\left[R_{\hU^\ell}
	(D+\Delta)-\zeta(s,d,\ell)-\epsilon(\ell,n)\right]\right\},
\end{eqnarray}
where $E_{\mbox{\tiny sp}}(R)$ is the sphere--packing exponent of the channel, i.e.,
\begin{equation}
	E_{\mbox{\tiny sp}}(R)=\sup_{Q_X}\inf_{\{Q_{Y|X}:~I_Q(X;Y)\le R\}}D(Q_{Y|X}\|P_{Y|X}|Q_X),
\end{equation}
with $I_Q(X;Y)$ denoting the mutual information induced by $Q_X\times Q_{Y|X}$.
\end{theorem}

As a simple conclusion from this theorem, we have that
\begin{equation}
	\liminf_{n\to\infty}\frac{1}{n}\log\left[ \mbox{Pr}\left\{\sum_{i=1}^n\rho(u_i,V_i)\ge nD\right\}\right]\ge
	-E_{\mbox{\tiny sp}}\left[R_{\hU^\ell}(D+0)-\zeta(s,d,\ell)\right],
\end{equation}
where
\begin{equation}
	R_{\hU^\ell}(D+0)\dfn\lim_{\Delta\downarrow 0} R_{\hU^\ell}(D+\Delta).
\end{equation}

\noindent
{\bf Discussion.}

First, observe that the ``extra capacity'' term, $\zeta(s,d,\ell)$, 
plays a role here too. This time, it appears in the form of an effective
rate reduction in the argument of the sphere--packing error exponent. But once again, 
if $\ell$ is very large while $s$ and $d$ are fixed, this term
becomes insignificant. In this case, as long as $D$ is smaller than $D_{\hU^\ell}(R_{\mbox{\tiny crit}})$,
$R_{\mbox{\tiny crit}}$ being the 
critical rate \cite{Gallager68} of the channel (and assuming $D_{\hU^\ell}(R_{\mbox{\tiny crit}})>0$),
the bound is asymptotically achievable using long blocks (of size $n\gg \ell$),
by rate--distortion coding (based on the type covering lemma) in the superalphabet of $\ell$--vectors, followed by channel coding, 
as in Csisz\'ar's works, \cite{Csiszar80} and \cite[Theorem 2]{Csiszar82}. 
As in those references, strictly speaking, this is not quite considered separate source-- and
channel coding, because there is a certain linkage between 
the channel code design and the source: Each type class of source sequences 
(in the level of $\ell$-blocks) is mapped into a channel sub--code at rate $R_{\hU^\ell}(D)$ (approximately), and the corresponding
channel codewords are of the type, $P_{\hX}$, that achieves the maximum sphere--packing exponent at that particular rate.

As a side remark,
speaking of Csisz\'ar's source--channel error exponents, \cite{Csiszar80} and \cite{Csiszar82}, 
it is interesting to relate Theorem 1 above to its purely probabilistic counterpart. The proof of Theorem 1 above is based on 
a change--of--measure argument. Since only the channel is probabilistic in our setting, the upper bound on the exponent includes 
only a channel--related term, which is the channel's sphere--packing exponent. Applying a similar line of thought in the purely
probabilistic case, we have to change measures for both the source and the channel, and so, 
we end up minimizing the sum of two divergence terms, i.e.,
\begin{equation}
	\min_{\{(Q_U,Q_{Y|X}):~R_{Q_U}(D)\ge I_Q(X;Y)\}}\{D(Q_U\|P_U)+D(Q_{Y|X}\|P_{Y|X}|Q_X)\},
\end{equation}
where $P_U$ is the memoryless source and $P_{Y|X}$ is the memoryless channel. This upper bound on the joint source--channel
exponent can be further upper bounded by arbitrarily selecting a positive real $R$ and arguing that
\begin{eqnarray}
	& &\min_{\{(Q_U,Q_{Y|X}):~R_{Q_U}(D)\ge I_Q(X;Y)\}}\{D(Q_U\|P_U)+D(Q_{Y|X}\|P_{Y|X}|Q_X)\}\nonumber\\
	&\le&\min_{\{(Q_U,Q_{Y|X}):~R_{Q_U}(D)\ge R\ge I_Q(X;Y)\}}\{D(Q_U\|P_U)+D(Q_{Y|X}\|P_{Y|X}|Q_X)\}\nonumber\\
	&=&\min_{\{(Q_U:~R_{Q_U}(D)\ge R\}}D(Q_U\|P_U)+\min_{\{Q_{Y|X}:~I_Q(X;Y)\le R\}}D(Q_{Y|X}\|P_{Y|X}|Q_X)\nonumber\\
	&\le&F(R,D,U)+E_{\mbox{\tiny sp}}(R),
\end{eqnarray}
where $F(R,D,U)$ is Marton's source coding exponent \cite{Marton74} for the memoryless source $U\sim P_U$.
Since this argument is applicable to any value of $R$, the tightest upper bound is obtained by minimizing over $R$,
namely, the resulting upper bound on the exponent is 
\begin{equation}
\min_R[F(R,D,U)+E_{\mbox{\tiny sp}}(R)], 
\end{equation}
which coincides with Csisz\'ar's
upper bound \cite[Theorem 4]{Csiszar82} to the best achievable excess--distortion exponent. 
This argument, however, is 
quite different from the one used in \cite{Csiszar82}, which in turn is based
on the list--decoding argument of Shannon, Gallager and Berlekamp \cite{SGB67}, that originally, sets the stage for
the straight--line bound \cite[Theorem 5.8.2]{Gallager68}.

Returning to the main topic of this work, the remaining part of this section is devoted to the proof of Theorem 1.\\
\noindent
{\em Proof of Theorem \ref{thm1}.}
Let $\Delta > 0$ be arbitrarily small and let $Q_{Y^n|X^n}(y^n|x^n)=
\prod_{i=1}^nQ_{Y|X}(y_i|x_i)$ be an auxiliary DMC
such that
\begin{equation}
	\label{spconstraint}
	\ell\cdot R_{\hat{U}^\ell}(D+\Delta)\ge \ell\cdot I_Q(\hat{X};Y)+\ell\zeta(s,d,\ell)+
	\ell\epsilon(\ell,n)=\ell[I_Q(\hat{X};Y)+\lambda],
\end{equation}
where $\lambda\dfn \zeta(s,d,\ell)+\epsilon(\ell,n)$, and
where the empirical channel input distribution $P_{\tX}$ is induced by the encoder output, $x^{n+d}$.
Since $Q_{Y|X}$ is assumed memoryless, we have the following relationship
between $I_Q(\tX;Y)$ and $I_Q(\tX^\ell;Y^\ell)$ ($\tX^\ell$ being the random vector induced by the empirical
distribution of $\ell$--blocks, extracted from $x^{n+d}$, assuming that $\ell$ divides $n+d$):
\begin{eqnarray}
	\label{IQ}
	\frac{I_Q(\tX^\ell;Y^\ell)}{\ell}&\le&\frac{1}{\ell}\sum_{j=1}^\ell I_Q(\tX_j;Y_j)\\
	&=&I(\tX_J;Y_J|J)\\
	&=&H(Y_J|J)-H(Y_J|\tX_J,J)\\
	&\le&H(Y_J)-H(Y_J|\tX_J,J)\\
	&=&H(Y_J)-H(Y_J|\tX_J)\\
	&=&I_Q(\tX_J;Y_J)\\
	&=&I_Q(\tX;Y),
\end{eqnarray}
where $\tX_j$ is the random variable derived from the the $j$--th marginal of $P_{\tX^\ell}$,
$j=1,2,\ldots,\ell$,
$J$ is an integer random variable, uniformly distributed over $\{1,2,\ldots,\ell\}$, and where we have used
the Markovity of the chain $J\to \hat{X}_J \to Y_J$, 
and the identities $\tX_J=\tX$,
$Y_J=Y$, which follow from the fact that $P_{\tX}=\frac{1}{\ell}\sum_{j=1}^\ell P_{\tX_j}=
P_{\tX_J}$. Therefore, if $Q$ satisfies (\ref{spconstraint}), it must also satisfy
\begin{equation}
	\label{spconstraint1}
	\ell\cdot R_{\hat{U}^\ell}(D+\Delta)\ge 
	I_Q(\tX^\ell;Y^\ell)+\ell\lambda.
\end{equation}
According to the above cited Theorem 1 of \cite{me14}, for such a channel, the expected distortion under $Q$, denoted
$D_Q$, must be lower bounded by
\begin{equation}
	D_Q\dfn\bE_Q\left\{\frac{1}{n}\sum_{i=1}^n\rho(u_i,V_i)\right\}\ge 
	 D_{\hU^\ell}\left(\frac{I_Q(\tX^\ell;Y^\ell)}{\ell}+
        \lambda\right)\ge
	D+\Delta.
\end{equation}
Denoting the event
$\calE=\{y^{n+d}:~\rho(u^n,v^n))> nD\}$, we have
\begin{eqnarray}
	D_Q&=&\frac{1}{n}\bE_Q\{\rho(u^n,V^n))\}\nonumber\\
	&=&\frac{1}{n}\bE_Q\{\rho(u^n,V^n)\cdot 1\{\calE\}\}+
	\frac{1}{n}\bE_Q\{\rho(u^n,V^n))\cdot 1\{\calE^{\mbox{\tiny c}}\}\}\nonumber\\
	&\le&\frac{1}{n}\cdot Q(\calE)\cdot n\rho_{\max}
	+\frac{1}{n}[1-Q(\calE)]\cdot nD\nonumber\\
	&=&[1-Q(\calE)]\cdot D+Q(\calE)\cdot \rho_{\max},
\end{eqnarray}
which implies that
\begin{equation}
	Q(\calE)\ge\frac{D_Q-D}{\rho_{\max}-D}\ge \frac{D+\Delta-D}{\rho_{\max}-D}=\frac{\Delta}{\rho_{\max}-D}.
\end{equation}
Now, for a given, arbitrarily small $\epsilon_0>0$, let us define
\begin{equation}
	\calT=\left\{y^{n+d}:~\sum_{i=1}^{n+d}\ln\frac{Q_{Y|X}(y_i|x_i)}{P_{Y|X}(y_i|x_i)}\le 
	(n+d)[D(Q_{Y|X}\|P_{Y|X}|P_{\tX})+\epsilon_0]\right\}.
\end{equation}
Now,
\begin{eqnarray}
	\mbox{Pr}\left\{\rho(u^n,V^n)\ge nD\right\}&=&\sum_{y^{n+d}\in\calE}
	P_{Y^{n+d}|X^{n+d}}(y^{n+d}|x^{n+d})\nonumber\\
	&\ge&\sum_{y^{n+d}\in\calE\cap\calT}
	P_{Y^{n+d}|X^{n+d}}(y^{n+d}|x^{n+d})\nonumber\\
	&=&\sum_{y^{n+d}\in\calE\cap\calT}
	Q_{Y^{n+d}|X^{n+d}}(y^{n+d}|x^{n+d})\cdot
	\exp\left\{-\sum_{i=1}^{n+d}\log\frac{Q_{Y|X}(y_i|x_i)}{P_{Y|X}(y_i|x_i)}\right\}\nonumber\\
	&\ge&\sum_{y^{n+d}\in\calE\cap\calT}
	Q_{Y^{n+d}|X^{n+d}}(y^{n+d}|x^{n+d})\cdot
	\exp\left\{-(n+d)[D(Q_{Y|X}\|P_{Y|X}|P_{\tX})+\epsilon_0]\right\}\nonumber\\
	&=&\exp\left\{-(n+d)[D(Q_{Y|X}\|P_{Y|X}|P_{\tX})+\epsilon_0]\right\}\cdot 
	Q(\calE\cap\calT)\nonumber\\
	&\ge&\exp\left\{-(n+d)[D(Q_{Y|X}\|P_{Y|X}|P_{\tX})+\epsilon_0]\right\}
	\cdot[Q(\calE)-Q(\calT^{\mbox{\tiny c}})]\nonumber\\
	&\ge&\exp\left\{-(n+d)[D(Q_{Y|X}\|P_{Y|X}|P_{\tX})+\epsilon_0]\right\}
	\cdot\left[\frac{\Delta}{\rho_{\max}-D}-o(n)\right],
\end{eqnarray}
where we have used the fact that $Q(\calT^{\mbox{\tiny c}})=o(n)$ for every $\epsilon_0> 0$, by the weak law of large numbers.
Since $Q_{Y|X}$ is an arbitrary channel that satisfies
(\ref{spconstraint}), we have
\begin{eqnarray}
	\mbox{Pr}\left\{\rho(u^n,V^n)\ge nD\right\}&\ge&\left[\frac{\Delta}{\rho_{\max}-D}-o(n)\right]\times\nonumber\\
	& &\exp\left\{-(n+d)\inf_{\{Q_{Y|X}:~I_Q(\hat{X};Y)\le R_{\hU^\ell}(D+\Delta)-
	\lambda\}}D(Q_{Y|X}\|P_{Y|X}|P_{\tX})+\epsilon_0]\right\}\nonumber\\
	&\ge&\left[\frac{\Delta}{\rho_{\max}-D}-o(n)\right]\cdot 
	\exp\left\{-(n+d)\left(E_{\mbox{\tiny sp}}[R_{\hU^\ell}
	(D+\Delta)-\lambda]+\epsilon_0\right)\right\},
\end{eqnarray}
which completes the proof of Theorem 1 by the arbitrariness of $\epsilon_0>0$.

\section{Side Information at the Decoder}
\label{wz}

Consider again the setting described in Section \ref{ncpf} and depicted in Fig.\ \ref{fig1}.
Our first result is the following:

\begin{theorem}
	\label{thm2}
	 Assume that $\rho_{\max}=\max_{u,v}\rho(u,v) < \infty$.
	For a given $u^n$, any finite--state encoder with $s_{\mbox{\tiny e}}$ states, 
	and any finite--state decoder with $s_{\mbox{\tiny d}}$ states, both having a period of length $\ell$ and an overall delay $d$,
	\begin{equation}
		\frac{1}{n}\sum_{i=1}^n\bE\{\rho(u_i,V_i)\}
		\ge D_{\hU^\ell|W^\ell}^{\mbox{\tiny
		WZ}}\left(C(\Gamma)+\frac{\log s_{\mbox{\tiny d}}}{\ell}+\frac{s_{\mbox{\tiny e}}\alpha^\ell\log \gamma}{\sqrt{n}}+o(n)
		\right)-\frac{\rho_{\max}d}{\ell}.
	\end{equation}
\end{theorem}

\noindent
{\bf Discussion.}

Similarly as in Theorem \ref{thm1}, the distortion lower bound depends on the source
sequence, $u^n$, only via its empirical distribution from the order that corresponds to the
period, $\ell$. We also observe that the 
numbers of states,
$s_{\mbox{\tiny e}}$, $s_{\mbox{\tiny d}}$
and the allowed delay, $d$, take parts in the lower bound 
in different ways. The first two play the role of
`excess capacity', whereas the latter serves in a term of distortion reduction. But this
difference is not really crucial, because excess rate and reduced distortion are two faces of
the same coin. Indeed, one of technical issues in the proof of Theorem \ref{thm2} below (which was 
not handled perfectly rigorously in the parallel derivation in \cite{me14}),
evolves around the following question: how can one
assess the effect of the decoder state contribution in estimating the source (and 
thereby reducing the distortion relative to the absence of the state), in order to bound the distortion in
terms of W-Z block code performance. In other words, the question is: how to obtain a lower
bound in terms of block codes, 
where no state carries information over from block to block? As will be seen in the proof below,
the idea is that since the decoder state cannot carry more than $\log
s_{\mbox{\tiny d}}$
information bits about the source, its effect cannot be better than that of adding an excess rate
of $\Delta R=(\log s_{\mbox{\tiny d}})/\ell$ to the corresponding W-Z encoder. 
This demonstrates clearly the point that distortion
reduction can be traded with excess rate. 

Another important observation is that our bound
depends very differently on $s_{\mbox{\tiny e}}$ and $s_{\mbox{\tiny d}}$. 
The dependence upon $s_{\mbox{\tiny e}}$, although linear rather than logarithmic,
is considerably weaker, as it vanishes as soon as the limit $n\to\infty$ is taken,
whereas the dependence on $s_{\mbox{\tiny d}}$ `survives' the limit $n\to\infty$ and vanishes only after the limit $\ell\to\infty$ is 
taken. In other words, at least as far as the distortion lower bound in concerned, the number of states of the decoder is a much more
important resource than the number of states of the encoder. We find this asymmetric behavior interesting and not trivial.
A partial intuitive explanation to this `discrimation' between encoder memory and decoder memory is the following: While at the encoder,
the state can only help to drive the input to achieve capacity, at the encoder, on the other hand, the state may play an important role in
helping to estimate the source by exploiting correlations with past source blocks, if exist.
It should also be pointed out that in terms of the dependence on $s_{\mbox{\tiny d}}$, our new lower bound is tighter than those of both
\cite{me14} and \cite{Ziv80} (in spite of the fact that we consider here a more general 
setting of side information at the decoder): the coefficient in front of
the extra-capacity term, $(\log s_{\mbox{\tiny d}})/\ell$, is reduced from 2 (in both \cite{me14} and \cite{Ziv80}) to 1 here.

For very large $\ell$ (compared to $d$
and $\log s_{\mbox{\tiny d}}$), the lower bound can be
asymptotically approached by separate source- and channel coding: W-Z coding in the superalphabet of $\ell$--vectors,
followed by channel coding. Specifically, given a long block of source sequence, $u^n$, of length $n \gg \alpha^\ell$, compute the
empirical distribution, $P_{\hU^\ell}$, and construct a W--Z code for the `source' $\{P_{\hU^\ell}(u^\ell),~u^\ell\in\calU^\ell\}$
and the side information channel $P_{W^\ell|\hU^\ell}$, for a distortion level tuned such that
$R_{\hU^\ell|W^\ell}^{\mbox{\tiny WZ}}(D) \le C(\Gamma)-2\epsilon$ for some prescribed, arbitrarily small $\epsilon>0$.
Append to the W-Z bitstream a header of 
length $\lceil\log(n/\ell+1)^{\alpha^\ell-1}\rceil$ in order to transmit to the decoder a description of the empirical distribution,
$\{P_{\hU^\ell}(u^\ell),~u^\ell\in\calU^\ell\}$. This is necessary 
for the receiver to apply the decoder corresponding to the encoder of that source.
If $n$ is large enough compared to $\alpha^\ell\log n$, then
\begin{equation}
	\frac{\lceil\log(n/\ell+1)^{\alpha^\ell-1}\rceil}{n}\le \frac{(\alpha^\ell-1)\log(n/\ell+1)+1}{n}\le\epsilon,
\end{equation}
and the total channel coding rate is below $C(\Gamma)-\epsilon$. By applying a good channel code for this rate,
the lower bound to the distortion is essentially achieved, similarly as in traditional separate source- and channel coding.

Block encoders of length $n$ can be thought of finite--state devices with delay $n$ and the same comment applies to block decoders.
While the numbers 
of states of this block encoder and decoder are much larger than $s_{\mbox{\tiny e}}$ and $s_{\mbox{\tiny d}}$, respectively,
the gaps between the converse and the achievability bounds shrinks in the asymptotic limit where $n$ tends to infinity and
then $\ell$ tends to infinity. The above described achievability scheme is similar in spirit to those of \cite{Ziv80} and \cite{Ziv84}.
Note that this separation--based scheme asymptotically meets the lower bound in spite of the fact the W-Z channel violates the Markov
structure of traditional communication system, $\bu\to\bx\to\by\to\bv$. This is coherent with the
analogous behavior in the purely probabilistic setting \cite{SVZ98}, and moreover, 
even if the DMC is
replaced by the Shannon channel.
In the last part of the next section, we will refer to this channel. 

Finally, on the basis of Theorem \ref{thm2}, and similarly as in Theorem \ref{thm1}, 
one can easily derive a lower bound on the
excess distortion probability for the case of decoder side information, considered here.
This time, however, the change of measures should involve, not only the main 
channel, $P_{Y|X}$, as before, but also the W-Z channel, $P_{W|\hU}$. The resulting exponential
lower bound would be of the form,
\begin{eqnarray}
\mbox{Pr}\left\{\sum_{i=1}^n\rho(u_i,V_i)\ge nD\right\}&\ge&
\sup_{\Delta> 0}\left[\frac{\Delta}{\rho_{\max}-D}-o(n)\right]\times\nonumber\\
& &\exp\bigg\{-(n+d)\max_{Q_{\hX}}\min\bigg[D(Q_{W|\hU}\|P_{W|\hU}|P_{\hU})+\nonumber\\
& &D(Q_{Y|\hX}\|P_{Y|\hX}|Q_{\hX})+2\epsilon_0\bigg]\bigg\},
\end{eqnarray}
where the minimum is over all pairs $\{(Q_{W|\hU},Q_{Y|\hX})\}$
such that
\begin{equation}
R_{\hU^\ell,Q_{W^\ell|\hU^\ell}}^{\mbox{\tiny WZ}}\left(D+\frac{\rho_{\max}d}{\ell}+\Delta\right)\ge 
	I_Q(\hX;Y)+\frac{\log s_{\mbox{\tiny d}}}{\ell}+\frac{s_{\mbox{\tiny e}}\alpha^\ell\log\gamma}{\sqrt{n}}+o(n),
\end{equation}
where $R_{\hU^\ell,Q_{W^\ell|\hU^\ell}}^{\mbox{\tiny WZ}}(\cdot)$ 
is the W-Z rate--distortion function of $P_{\hU^\ell}$ with the W-Z channel, $Q_{W^\ell|\hU^\ell}$.
Similarly as before, the bound is asymptotically achievable following the same line of thought as
in \cite{Csiszar80} and \cite{Csiszar82}, for 
$D < D_{\hU^\ell|W^\ell}^{\mbox{\tiny WZ}}(R_{\mbox{\tiny crit}})$, where $R_{\mbox{\tiny crit}}$
is the critical rate of the channel, provided that
$D_{\hU^\ell|W^\ell}^{\mbox{\tiny WZ}}(R_{\mbox{\tiny crit}})>0$.

The remaining part of this section is devoted to the proof of Theorem \ref{thm2}.\\

\noindent
{\em Proof of Theorem \ref{thm2}.}
Owing to the encoder--decoder model (\ref{fsme}), (\ref{fsmd}), it is clear that
$x_{i\ell+1}^{i\ell+\ell}$ is a deterministic function of
$z_{i\ell+1}^{\mbox{\tiny e}}$ and $u_{i\ell+1}^{i\ell+\ell}$, as
\begin{eqnarray}
	x_{i\ell+1}&=&f_1(u_{i\ell+1},z_{i\ell+1}^{\mbox{\tiny e}}),\nonumber\\
	x_{i\ell+2}&=&f_2(u_{i\ell+2},z_{i\ell+2}^{\mbox{\tiny e}})=
	f_2(u_{i\ell+2},g_1(u_{i\ell+1},z_{i\ell+1}^{\mbox{\tiny e}}))\nonumber\\
	&\ldots&\nonumber\\
	x_{i\ell+\ell}&=&f_0(u_{i\ell+\ell},z_{i\ell+\ell}^{\mbox{\tiny e}}).
\end{eqnarray}
Accordingly, we denote
\begin{equation}
	x_{i\ell+1}^{i\ell+\ell}=q(z_{i\ell+1}^{\mbox{\tiny
e}},u_{i\ell+1}^{i\ell+\ell}),
\end{equation}
where $q:\calZ^{\mbox{\tiny e}}\times\calU^\ell\to\calX^\ell$.
Likewise, $v_{i\ell+1}^{i\ell+\ell-d}$ is a deterministic function of
$z_{i\ell+1}^{\mbox{\tiny d}}$, $y_{i\ell+1}^{i\ell+\ell}$ 
and $w_{i\ell+1}^{i\ell+\ell}$, as
\begin{eqnarray}
	v_{i\ell+1}&=&f_{d+1}'(w_{i\ell+d+1},y_{i\ell+d+1},z_{i\ell+d+1}^{\mbox{\tiny d}}),~~~~
	z_{i\ell+d+1}^{\mbox{\tiny d}}~\mbox{being a function of
$w_{i\ell+1}^{i\ell+d}$, $y_{i\ell+1}^{i\ell+d}$ and $z_{i\ell+1}^{\mbox{\tiny d}}$}\nonumber\\
	v_{i\ell+2}&=&f_{d+2}'(w_{i\ell+d+2},y_{i\ell+d+2},z_{i\ell+d+2}^{\mbox{\tiny d}})\nonumber\\
	&\ldots&\nonumber\\
	v_{i\ell+\ell-d}&=&f_0'(w_{i\ell+\ell},y_{i\ell+\ell},z_{i\ell+\ell}^{\mbox{\tiny d}}).
\end{eqnarray}
Accordingly, we denote
\begin{equation}
	v_{i\ell+1}^{i\ell+\ell-d}=m(z_{i\ell+1}^{\mbox{\tiny
d}},w_{i\ell+1}^{i\ell+\ell},y_{i\ell+1}^{i\ell+\ell}),
\end{equation}
where $m:\calZ^{\mbox{\tiny d}}\times\calW^\ell\times\calY^\ell\to\calV^{\ell-d}$.
The proof of the theorem is based on deriving both a lower bound and an upper 
bound to the expected empirical conditional mutual information,
$\bE\{I(\hU^\ell;\hY^\ell)\}$, which applies to
any finite--state encoder and any finite--state decoder with an overall delay $d$. 

As for an upper bound, we have following.
\begin{eqnarray}
	\label{ub}
	\bE\{I(\hU^\ell;\hY^\ell)\}
	&\le&\bE\{I(\hU^\ell,\hZ^{\mbox{\tiny e}};\hY^\ell)\}\nonumber\\
	&=&\bE\{H(\hY^\ell)\}
-\bE\{H(\hY^\ell|\hU^\ell,\hZ^{\mbox{\tiny
e}})\}\nonumber\\
	&\le&H(Y^\ell)-\bE\{H(\hY^\ell|\hU^\ell,\hZ^{\mbox{\tiny e}})\},
\end{eqnarray}
where the last inequality follows from the concavity of the entropy function.
In Appendix A, we prove the following inequality:
\begin{equation}
	\label{expectedempiricalentropy}
	\bE\{H(\hY^\ell|\hU^\ell,\hZ^{\mbox{\tiny e}})\}\ge H(Y^\ell|\hX^\ell)-\ell\cdot\Delta_1(s_{\mbox{\tiny e}},\ell,n),
\end{equation}
where
\begin{equation}
       \label{delta1}
       \Delta_1(s_{\mbox{\tiny e}},\ell,n)\dfn
        \frac{s_{\mbox{\tiny e}}\alpha^\ell\log \gamma}{\sqrt{n}}+
o\left(\frac{1}{\sqrt{n}}\right).
\end{equation}
Combining this with eq.\ (\ref{ub}), we get
\begin{eqnarray}
	\label{cap}
	\bE\{I(\hU^\ell;\hY^\ell)\}&\le&H(Y^\ell)-H(Y^\ell|\hX^\ell)+\ell\cdot\Delta_1(s_{\mbox{\tiny e}},\ell,n)\nonumber\\
	&=&I(\hX^\ell;Y^\ell)+\ell\cdot\Delta_1(s_{\mbox{\tiny e}},\ell,n)\nonumber\\
	&\le&\ell\cdot[C(\Gamma)+\Delta_1(s_{\mbox{\tiny e}},\ell,n)].
\end{eqnarray}
Note in passing that in the last step of (\ref{ub}) 
one could use a tighter upper bound: instead of maximizing over all
$\{P_{X^\ell}\}$ that comply with the transmission cost constraint, we could have also maximized over
all $\{P_{X^\ell}\}$ that maintain the same empirical single--letter 
marginal, $P_{\hX}$ (refer to eq.\ (\ref{IQ})).
While this fact is immaterial for the
expected distortion lower bound, it will be important when it comes to the lower bound on the excess--distortion
probability. 

To derive a lower bound to $\bE\{I(\hU^\ell;\hY^\ell)\}$,
we first underestimate
$I(\hU^\ell;\hY^\ell)$ without
taking the expectation.
\begin{eqnarray}
	\label{ratelb}
	I(\hU^\ell,\hY^\ell)
	&=&I(\hU^\ell,W^\ell;\hY^\ell)\nonumber\\
	&\ge&I(\hU^\ell,W^\ell;\hY^\ell)-
I(W^\ell;\hY^\ell)\nonumber\\
	&=&I(\hU^\ell;\hY^\ell|W^\ell)\nonumber\\
	&\ge&\ell\cdot R_{\hU^\ell|W^\ell}^{\mbox{\tiny
WZ}}[\Delta(\hU^\ell|\hY^\ell,W^\ell)/\ell],
\end{eqnarray}
where the underlying joint distribution of $(\hU^\ell,\hY^\ell,W^\ell)$ is assumed 
$P_{\hU^\ell,\hY^\ell}\times P_{W^\ell|\hU^\ell}$.
The first equality follows from the fact that 
$W^\ell\to \hU^\ell\to \hY^\ell$ forms a Markov chain.
Consider now the joint distribution 
$P_{\hU^\ell W^\ell\hY^\ell\hZ^{\mbox{\tiny d}}}=P_{\hU^\ell
\hY^\ell\hZ^{\mbox{\tiny d}}}\times P_{W^\ell|\hU^\ell}$, 
where $\hZ^{\mbox{\tiny d}}$ designates the decoder state whose
alphabet size is $s_{\mbox{\tiny d}}$. 
We argue that no matter what this distribution may be, the best 
resulting distortion in estimating
$\hU^\ell$ from $(W^\ell,\hY^\ell,\hZ^{\mbox{\tiny d}})$, that is,
$\Delta(\hU^\ell|W^\ell,\hY^\ell,\hZ^{\mbox{\tiny d}})$, 
cannot be better than the minimum achievable
distortion when $\hZ^{\mbox{\tiny d}}$ is replaced by another random variable,
$Z_*^{\mbox{\tiny d}}$, 
of the same alphabet size $s_{\mbox{\tiny d}}$, that is given by a deterministic function of
$\hU^\ell$ (see also \cite[Proof of the converse to Theorem 6]{LMW14} for the
use of a similar idea, albeit in a very different context). Indeed,
\begin{eqnarray}
	\Delta(\hU^\ell|W^\ell,\hY^\ell,\hZ^{\mbox{\tiny d}})&=&
	\min_G\sum_{u^\ell}P_{\hU^\ell}(u^\ell)\sum_{z^{\mbox{\tiny
d}}}P_{\hZ^{\mbox{\tiny d}}|\hU^\ell}(z^{\mbox{\tiny d}}|u^\ell)\times\nonumber\\
	& &\sum_{w^\ell,y^\ell}P_{W^\ell\hY^\ell|\hU^\ell\hZ^{\mbox{\tiny d}}}
(w^\ell,y^\ell|u^\ell,z^{\mbox{\tiny
d}})\rho(u^\ell,G(w^\ell,y^\ell,z^{\mbox{\tiny d}}))
\end{eqnarray}
is minimized by the conditional distribution, $P_{\hZ^{\mbox{\tiny d}}|\hU^\ell}$, that puts all its mass on
\begin{equation}
	z_*^{\mbox{\tiny d}}(u^\ell)=\mbox{arg}\min_{z^{\mbox{\tiny d}}}
c(u^\ell,z^{\mbox{\tiny d}}),
\end{equation}
where 
\begin{equation}
	c(u^\ell,z^{\mbox{\tiny
d}})\dfn\sum_{w^\ell,y^\ell}P_{W^\ell\hY^\ell|\hU^\ell\hZ^{\mbox{\tiny
d}}}(w^\ell,y^\ell|u^\ell,z^{\mbox{\tiny d}})
	\rho(u^\ell,G(w^\ell,y^\ell,z^{\mbox{\tiny d}})).
\end{equation}
Since $Z_*^{\mbox{\tiny d}}$ is a deterministic function of $\hU^\ell$, it is available to the encoder.
Consider now a coding scheme that transmits 
$I(\hU^\ell;\hY^\ell|W^\ell)$ bits per $\ell$--vector using a Wyner--Ziv code
(with $W^\ell$ serving as side information at the decoder),
plus additional $\log s_{\mbox{\tiny d}}$ bits to transmit $Z_*^{\mbox{\tiny
d}}$ as additional information on $\hU^\ell$. 
The overall code--length
of $I(\hU^\ell;\hY^\ell|W^\ell)+\log s_{\mbox{\tiny d}}$ bits cannot be smaller than that of the best 
Wyner--Ziv code that makes $(W^\ell,\hY^\ell,Z_*^{\mbox{\tiny d}})$
available to the decoder, with an extra rate of $\log s_{\mbox{\tiny d}}$ bits,
as the latter can potentially exploit the statistical dependence between
$\hU^\ell$ and $Z_*^{\mbox{\tiny d}}$.
In other words,
the point $((I(\hU^\ell;\hY^\ell|W^\ell)+\log s_{\mbox{\tiny
d}})/\ell,\Delta(\hU^\ell|W^\ell,\hY^\ell,Z_*^{\mbox{\tiny d}})/\ell)$ 
is an achievable point on
the rate--distortion plane, which implies that
\begin{equation}
	I(\hU^\ell;\hY^\ell|W^\ell)+\log s_{\mbox{\tiny d}}\ge \ell\cdot R_{\hU^\ell|W^\ell}^{\mbox{\tiny WZ}}
	[\Delta(\hU^\ell|W^\ell,\hY^\ell,Z_*^{\mbox{\tiny d}})/\ell]\ge \ell\cdot
	R_{\hU^\ell|W^\ell}^{\mbox{\tiny
WZ}}[\Delta(\hU^\ell|W^\ell,\hY^\ell,\hZ^{\mbox{\tiny d}})/\ell],
\end{equation}
where the last equality follows 
from $\Delta(\hU^\ell|W^\ell,\hY^\ell,Z_*^{\mbox{\tiny d}})\le
\Delta(\hU^\ell|W^\ell,\hY^\ell,\hZ^{\mbox{\tiny d}})$ and the
non--increasing monotonicity of the Wyner--Ziv rate--distortion function.

Let us now focus on the quantity
$\Delta(\hU^\ell|W^\ell,\hY^\ell,\hZ^{\mbox{\tiny d}})/\ell$:
\begin{eqnarray}
	\label{distortionub}
	\frac{1}{\ell}\Delta(\hU^\ell|W^\ell,\hY^\ell,\hZ^{\mbox{\tiny d}})
&=&\min_G\frac{1}{\ell}\sum_{u^\ell,w^\ell,y^\ell,z^{\mbox{\tiny d}}}
	P_{\hU^\ell W^\ell\hY^\ell \hZ^{\mbox{\tiny
d}}}(u^\ell,w^\ell,y^\ell,z^{\mbox{\tiny
d}})\rho(u^\ell,G(w^\ell,y^\ell,z^{\mbox{\tiny d}}))\nonumber\\
	&\le&\frac{1}{\ell}\sum_{u^\ell,w^\ell,y^\ell,z^{\mbox{\tiny d}}}
	P_{\hU^\ell W^\ell\hY^\ell \hZ^{\mbox{\tiny
d}}}(u^\ell,w^\ell,y^\ell,z^{\mbox{\tiny
d}})[\rho(u^{\ell-d},m(z^{\mbox{\tiny d}},w^\ell,y^\ell))+\rho_{\max}\cdot d]\nonumber\\
	&=&\bE\left\{\frac{1}{n}\sum_{i=0}^{n/\ell-1}\left[\sum_{\tau=1}^{\ell-d}
	\rho(u_{i\ell+\tau},f_{\tau+d}'(W_{i\ell+\tau+d},y_{i\ell+\tau+d},z_{i\ell+\tau+d}^{\mbox{\tiny
d}}))+\rho_{\max}\cdot d\right]\right\}\nonumber\\
	&\le&\bE_W\left\{\frac{1}{n}\sum_{t=1}^{n}
	\rho(u_t,V_t)\right\}+\frac{\rho_{\max}\cdot d}{\ell},
\end{eqnarray}
where $\bE_W$ denotes expectation w.r.t.\ $W^n$ only.
The first inequality follows from the definition of $G$ as the optimal
estimator and the fact that, due to the delay, the estimates of the last $d$
source symbols are not yet available in the current $\ell$--block, but their
distortions cannot exceed $\rho_{\max}$. The subsequent equality is by
the definitions of all the ingredients involved, and by passing from empirical distributions back to time averages.
Thus,
\begin{equation}
	I(\hU^\ell;\hY^\ell|W^\ell)+\log s_{\mbox{\tiny d}}\ge\ell\cdot 
	R_{\hU^\ell|W^\ell}^{\mbox{\tiny WZ}}\left(\bE_W\left\{\frac{1}{n}\sum_{i=1}^{n}
	\rho(u_i,V_i)\right\}+\frac{\rho_{\max}\cdot d}{\ell}\right),
\end{equation}
and consequently, following the second to the last line of (\ref{ratelb}), we have
\begin{eqnarray}
	I(\hU^\ell,\hY^\ell)
&\ge& I(\hU^\ell;\hY^\ell|W^\ell)\nonumber\\
&=& I(\hU^\ell;\hY^\ell|W^\ell)+\log s_{\mbox{\tiny
d}}-\log s_{\mbox{\tiny d}}\nonumber\\
&\ge&\ell\cdot 
        R_{\hU^\ell|W^\ell}^{\mbox{\tiny
WZ}}\left(\bE_W\left\{\frac{1}{n}\sum_{i=1}^{n}
        \rho(u_i,V_i)\right\}+\frac{\rho_{\max}\cdot d}{\ell}\right)-
\log s_{\mbox{\tiny d}}.
\end{eqnarray}
Finally, combining this with (\ref{cap}) and taking the expectation 
w.r.t.\ the randomness of the channels, we have
\begin{eqnarray}
	\ell\cdot C(\Gamma)+\log s_{\mbox{\tiny d}}+\ell\cdot\left[
        \frac{s_{\mbox{\tiny e}}\alpha^\ell\log \gamma}{\sqrt{n}}+
	+o\left(\frac{1}{\sqrt{n}}\right)\right]
	&\ge&\bE\left\{R_{\hU^\ell|W^\ell}^{\mbox{\tiny WZ}}\left(\bE_W\left\{\frac{1}{n}\sum_{i=1}^{n}
\rho(u_i,V_i)\right\}+\frac{\rho_{\max}\cdot d}{\ell}\right)\right\}\nonumber\\
	&\ge&R_{\hU^\ell|W^\ell}^{\mbox{\tiny WZ}}\left(\frac{1}{n}\sum_{i=1}^{n}\bE\left\{\rho(u_i,V_i)\right\}+
	\frac{\rho_{\max}\cdot d}{\ell}\right),
\end{eqnarray}
where in the last inequality we have used Jensen's inequality and the convexity of the W-Z rate--distortion function
\cite[Lemma 15.9.1]{CT06}. The
assertion of Theorem \ref{thm2} now follows immediately.

\section{Variations, Modifications and Extensions}
\label{ext}

In this section, we outline a few variants of our main results that require a few changes in the
model considered. We discuss the following modifications: (i) the additional constraint of common reconstruction, (ii)
the case where side information is available to the encoder too, and (iii) channel state information at the encoder.
In all these cases, we discuss the changes needed in proof of Theorem \ref{thm2}, and one can also
obtain a lower bound to the excess distortion probability by applying the appropriate change of measures, following
the same ideas as in the proof of Theorem \ref{thm1}, and as discussed after Theorem \ref{thm2}.

\subsection{Common Reconstruction}

In \cite{Steinberg09}, Steinberg studied a version of the W-Z problem \cite{WZ76}, where there is an additional constraint that
the encoder would be capable of generating an {\em exact} copy of the reconstruction sequence 
to be generated by the decoder, with motivation in medical imaging, etc. In the ordinary W-Z setting, this is not the case since
the reconstruction depends on the side information, which is not available to the encoder. Steinberg's solution to the W-Z problem
with common reconstruction is very similar to the solution of the regular W-Z problem: the only difference is that the estimator
at the decoder is allowed to be a function of the compressed representation only, rather than being a function of both the
compressed representation and the side information vector. 
In other words, in Steinberg's scheme, the side information serves the decoder
only for the purpose of binning, and not for both binning and estimation, as in the classical W-Z achievability scheme.
For a pair of memoryless correlated sources, $\{(U_i,W_i)\}$, Steinberg's coding theorem \cite[Theorem 1]{Steinberg09} 
for coding under the common reconstruction
constraint, asserts that the corresponding rate--distortion function is given by
\begin{equation}
	R_{U|W}^{\mbox{\tiny WZ,cr}}(D)=\min I(U;V|W)\equiv\min\{I(U;V)-I(W;V)\},
\end{equation}
where the minimum is over all conditional distributions, $\{P_{V|U}\}$, such that $V\to U\to W$ is a
Markov chain and $\bE\{\rho(U,V)\}\le D$.

Equipped with this background, we can impose the common reconstruction constraint in our setting too, provided that the
model of the finite--state decoder is somewhat altered: 
Instead of feeding the finite--state decoder sequentially by $\{(w_i,y_i)\}$,
as before, we now feed it by a single sequence, 
$\{r_i\}$, where $r^n=(r_1,\ldots,r_n)$ is a deterministic function of $u^n$, which
with very high probability (for large $n$), can be reconstructed faithfully at the decoder as a function of $(w^n,y^n)$.

The modifications needed in the proof of Theorem \ref{thm2} are in two places only: 
The first modification is that in the last line of eq.\ (\ref{ratelb}), 
$R_{\hU^\ell|W^\ell}^{\mbox{\tiny WZ}}[\Delta(\hU^\ell|\hY^\ell,W^\ell)/\ell]$ should be replaced by
$R_{\hU^\ell|W^\ell}^{\mbox{\tiny WZ,cr}}[\bE\{\rho(\hU^\ell,\hV^\ell)\}/\ell]$, where following \cite{Steinberg09},
$R_{\hU^\ell|W^\ell}^{\mbox{\tiny WZ,cr}}(D)$ is defined according to
\begin{equation}
	R_{\hU^\ell|W^\ell}^{\mbox{\tiny WZ,cr}}(D)=\frac{1}{\ell}\min I(\hU^\ell;V^\ell|W^\ell),
\end{equation}
where the minimum is over all $\{P_{\hV^\ell|\hU^\ell}\}$ such that $\hV^\ell\to\hU^\ell\to W^\ell$ is a Markov chain and
$\bE\{\rho(\hU^\ell,\hV^\ell)\le \ell\cdot D$. The second modification is in eq.\ (\ref{distortionub}), 
where $G(w^\ell,y^\ell,z^{\mbox{\tiny d}})$, $m(z^{\mbox{\tiny d}},w^\ell,y^\ell)$ 
and
$f_{\tau+d}(W_{i\ell+\tau+d},y_{i\ell+\tau+d},z_{i\ell+\tau+d}^{\mbox{\tiny
d}})$
should be replaced by
$G(r^\ell,z^{\mbox{\tiny d}})$, $m(z^{\mbox{\tiny d}},r^\ell)$, and 
$f_{\tau+d}(r_{i\ell+\tau+d},z_{i\ell+\tau+d}^{\mbox{\tiny d}})$, respectively.

The achievability is based on source coding using 
Steinberg's coding scheme \cite{Steinberg09}, followed by a capacity--achieving channel code.

\subsection{Side Information at Both Ends}

So far, we have considered the case where the side information is available at the decoder only.
On the face of it, one might argue that there is 
no much point to address the case where side information is available to both encoder and decoder, because it is much easier and it
can even be viewed as a special case of side information at the decoder only (simply by redefining $\{(u_i,w_i)\}$ as the ``source'').
Nevertheless, we mention the case of two--sided side information for two reasons:
\begin{enumerate}
	\item We can allow $\bw$ to be an individual sequence too, in addition to $\bu$, as opposed to our assumption so far that
		it is generated by a DMC fed by $\bu$.
	\item We can derive more explicit lower bounds to the distortion. These bounds automatically apply also to the case
		where the side information is available to the decoder only, although they might not be tight for that case.
\end{enumerate}

In the purely probabilistic setting, the rate--distortion function in the presence of side information at both ends is given by
the so called conditional rate--distortion function.
As discussed in \cite{Zamir96}, the only difference between the W-Z rate--distortion function and the conditional rate--distortion
function is that in the former, there is the constraint of the Markov structure, whereas the conditional rate--distortion function
this constraint is dropped. The proof of Theorem \ref{thm2} 
can easily be altered to incorporate two--sided availability of side information, with
both $\bu$ and $\bw$ being deterministic sequences. The only modification needed is in eq.\ (\ref{ratelb}), which will now read as
follows:
\begin{equation}
	I(\hU^\ell;\hY^\ell)\ge \ell\cdot R_{\hU^\ell}[\Delta(\hU^\ell|\hY^\ell)/\ell]
	\ge \ell\cdot R_{\hU^\ell|\hW^\ell}[\Delta(\hU^\ell|\hY^\ell,\hW^\ell)/\ell],
\end{equation}
where the second inequality is due to the fact that ignoring side information at both ends cannot be better than
using it optimally. Note that we have also replaced $W^\ell$ by $\hW^\ell$, to account for the fact that we allow it to be
a deterministic sequence too, as mentioned before. Obviously, achievability is by conditional rate--distortion coding followed
by capacity--achieving channel coding.

For the purpose of the lower bound to the distortion, we can further lower bound the empirical conditional
rate--distortion function as follows in the spirit of the conditional Shannon lower bound \cite{Gray73}.
First, we can represent it as
\begin{equation}
	\label{cslb}
	R_{\hU^\ell|\hW^\ell}(D)=H(\hU^\ell|\hW^\ell)-\max_{\{P_{\hV^\ell|\hU^\ell\hW^\ell}:~\bE\{\rho(\hU^\ell,\hV^\ell)\le\ell D\}}
	H(\hU^\ell|\hW^\ell,\hV^\ell).
\end{equation}
Now, the first term, $H(\hU^\ell|\hW^\ell)$, can be further lower bounded (within an asymptotically 
negligible term) in terms of the conditional Lempel-Ziv
code--length function of $u^n$ given $w^n$ (as side information at both ends), as defined in \cite{Ziv85}.
Specifically, the following inequality is derived in \cite[eq.\ (17)]{me00}:
\begin{equation}
	H(\hU^\ell|\hW^\ell)\ge \frac{\ell}{n}\sum_{j=1}^{c(w^n)}
	[c_j(u^n|w^n)+q^2]\log\frac{c_j(u^n|w^n)}{4q^2},
\end{equation}
where $q$ is a constant that depends only on $\ell$ and on the sizes of $\calU$ and $\calW$,
$c(w^n)$ denotes the number of distinct phrases of $w^n$ that appear in joint incremental parsing \cite{ZL78} of $(u^n,w^n)$,
and $c_j(u^n|w^n)$ is
the number of distinct phrases of $u^n$ that appear jointly with the $j$--th distinct phrase of $w^n$.
Similarly as in \cite[Theorem 2]{me14}, for the case where $\calU=\calV=\{0,1,\ldots,\alpha-1\}$ 
and a difference distortion measure, $\rho(u,v)=\varrho(u-v)$ (where the subtraction is defined
modulo $\alpha$),
the second, subtracted term of (\ref{cslb}), can be easily upper bounded by the constrained maximum entropy function,
i.e., 
\begin{equation}
\max_{\{P_{\hV^\ell|\hU^\ell\hW^\ell}:~\bE\{\rho(\hU^\ell,\hV^\ell)\le\ell D\}}
	H(\hU^\ell|\hW^\ell,\hV^\ell)\le\ell\cdot\Phi(D),
\end{equation}
where
\begin{equation}
	\label{phi}
	\Phi(D)=\sup_{\theta\ge 0}\left[\theta D+\log\left(\sum_{u\in\calU}2^{-\theta\varrho(u)}\right)\right],
\end{equation}
as can easily be shown by the standard solution to the problem of maximum entropy under a moment constraint.
It then follows that the expected distortion is lower bounded in terms of the inverse function, $\Psi=\Phi^{-1}$, computed at the
difference,
$$\frac{1}{n}\sum_{j=1}^{c(w^n)}c_j(u^n|w^n)\log c_j(u^n|w^n)-C(\Gamma)-\eta(s_{\mbox{\tiny e}},s_{\mbox{\tiny e}},d,\ell,n)$$
where $\eta(s_{\mbox{\tiny e}},s_{\mbox{\tiny d}},d,\ell,n)$ accounts for all resulting redundancy terms 
(similarly as in \cite[proof of Theorem 2]{me14}). Specifically, the inverse function, $\Psi$, is
given by (see Appendix B)
\begin{equation}
	\label{psi}
	\Psi(R)=\inf_{\vartheta\ge 0}\vartheta\cdot\left[R-\log\left(\sum_{u\in\calU}2^{-\varrho(u)/\vartheta}\right)\right],
\end{equation}
and so,
\begin{eqnarray}
	\frac{1}{n}\sum_{i=1}^n\bE\{\rho(u_i,V_i)\}&\ge&\sup_{\vartheta\ge 0}\vartheta\cdot\bigg[
		\frac{1}{n}\sum_{j=1}^{c(w^n)}c_j(u^n|w^n)\log c_j(u^n|w^n)-
		C(\Gamma)-\nonumber\\
		& &\eta(s_{\mbox{\tiny e}},s_{\mbox{\tiny d}},d,\ell,n)-
		\log\bigg(\sum_{u\in\calU}2^{-\varrho(u)/\vartheta}\bigg)\bigg]-\frac{\rho_{\max}d}{\ell}.
\end{eqnarray}
Of course, instead of maximizing over $\vartheta$, one may select any arbitrary positive $\vartheta$ and thereby obtain a
a valid lower bound, albeit not as tight.
We observe that whenever
the conditional LZ complexity, $\frac{1}{n}\sum_{j=1}^{c(w^n)}c_j(u^n|w^n)\log c_j(u^n|w^n)$, exceeds
the channel capacity (plus the redundancy terms), the expected distortion has a non--trivial, strictly positive lower bound.

The advantage of the use of the conditional LZ complexity is that it is easier to calculate than the $\ell$--th order empirical
entropy, especially when $\ell$ is large, as the super-alphabet size grows exponentially with $\ell$. In other words,
we sacrifice tightness to a certain extent, at the benefit of facilitating the calculation of the bound.

\subsection{Channel--State Information at the Encoder}

The last extension that we discuss in this work is from the DMC, $P_{Y|X}$, to a state--dependent channel model, 
$P_{Y^n|X^nS^n}=[P_{Y|XS}]^n$, where the
state sequence, $s^n$, which is governed by a discrete 
memoryless source, $P_{S^n}=[P_S]^n$ of alphabet $\calS$ of size $\sigma$, 
is fed into the encoder as well. Specifically, the finite--state encoder
is now described by the following set of recursive equations:
\begin{eqnarray}
        t&=&i~\mbox{mod}~\ell\\
x_i&=&f_t(u_i,s_i,z_i^{\mbox{\tiny e}}),\\
z_{i+1}^{\mbox{\tiny e}}&=&g_t(u_i,s_i,z_i^{\mbox{\tiny e}}).
\end{eqnarray}
Since $\{s_i\}$ are fed sequentially into the 
encoder, this clearly falls in the category of causal state information \cite{Shannon58}.
Here, eq.\ (\ref{ub}) in the proof of Theorem \ref{thm2} should be modified as follows.
Let $P_{\hU^\ell S^\ell}=P_{\hU^\ell}\times P_{S^\ell}=P_{\hU^\ell}\times[P_S]^\ell$.
We also define
\begin{equation}
	\calP_\ell(\Gamma)=\{(P_{B},L):~X^\ell=L(B,S^\ell),
        \bE_{P_{X^\ell}}\phi(X^\ell)\le\ell\Gamma\},
\end{equation}
where $B$ is an auxiliary random variable whose alphabet size need not be larger than 
$\min\{\beta^\ell-1,\sigma^\ell+1,\gamma^\ell\}$, and is independent of $S^\ell$ \cite[Theorem 7.2]{EGK11},
and where $\phi(\hX^\ell)$ is the additive extension of the single--letter transmission cost function,
that is, $\phi(\hX^\ell)=\sum_{i=1}^\ell\phi(\hX_i)$. Then,
\begin{eqnarray}
	\bE\{I(\hU^\ell;\hY^\ell)\}
	&=&\bE\{H(\hY^\ell)\}
-\bE\{H(\hY^\ell|\hU^\ell)\}\\
        &\le&H(Y^\ell)-\bE\{H(\hY^\ell|\hU^\ell)\}\\
	&\le&H(Y^\ell)-H(Y^\ell|\hU^\ell)
	+\ell\cdot o(n)\\
	&=&I(\hU^\ell;Y^\ell)+
\ell\cdot o(n)\\
	&\le&\max_{(P_B,L)\in\calP_\ell(\Gamma)}
	I(B;Y^\ell)+
\ell\cdot o(n)\\
	&\le&\ell\cdot \left[C_{\mbox{\tiny S}}(\Gamma)
+o(n)\right],
\end{eqnarray}
where $C_{\mbox{\tiny S}}(\Gamma)$ is the capacity of the Shannon channel 
of causal channel state information \cite{Shannon58}
with average transmission 
cost limited by $\Gamma$ \cite[eqs.\ (3.9), (3.33)]{KSM07}.
The four inequalities of this chain are explained as
follows. The first inequality is due to the
concavity of the entropy as a functional of the underlying distribution.
The second one is obtained
by the weak law of large numbers, which guarantees that
$P_{\hY^\ell|X^\ell S^\ell}\to P_{Y^\ell|X^\ell S^\ell}$, in probability, as $n\to\infty$ (for fixed $\ell$), and
therefore, 
$$P_{\hY^\ell|\hU^\ell}(y^\ell|u^\ell)=
\sum_{s^\ell,z^{\mbox{\tiny e}}}P_{\hZ^{\mbox{\tiny e}}|\hU^\ell}(z^{\mbox{\tiny e}}|u^\ell)
P_{S^\ell}(s^\ell)P_{\hY^\ell|X^\ell S^\ell}(y^\ell|q[u^\ell,s^\ell,z^{\mbox{\tiny e}}],s^\ell)$$
tends to $P_{Y^\ell|\hU^\ell}(y^\ell|u^\ell)$ in probability for all $u^\ell\in\calU^\ell,~y^\ell\in\calY^\ell$.
The third inequality
is due to the fact that $\hU^\ell$ (like any feasible $B$) is independent of $S^\ell$ and 
since the 
reference encoding function, $q$, is assumed to comply with the transmission cost constraint. Note that
the additional dependence of $x^\ell$ upon $z^{\mbox{\tiny e}}$ can be viewed as 
randomized encoding according to
$$P(x^\ell|u^\ell,s^\ell)=\sum_{z^{\mbox{\tiny e}}}
P_{\hZ^{\mbox{\tiny e}}|\hU^\ell}(z^{\mbox{\tiny e}}|u^\ell)\cdot 1\{x^\ell=q[u^\ell,s^\ell,z^{\mbox{\tiny e}}]\},$$
that cannot 
improve capacity. Finally,
the last inequality is due to the fact that the multi-letter extension of the capacity formula cannot improve
on the single--letter version, as can easily been seen by comparing the highest
rate achievable by multi--letter random coding
(over the superalphabet of $\ell$--vectors) to the converse bound on the highest achievable rate, which is given by
the single--letter formula.

Finally, note that if the allowed delay, $d$, exceeds the period $\ell$, and $s_{\mbox{\tiny e}}\ge\alpha^\ell$, 
then the encoder can afford to wait until the
end of the $\ell$--block (and store it as its state) before beginning to encode. In this case, the more relevant
capacity formula is the one of non--causal state information \cite{GP80}, which is, in general, larger than $C_S(\Gamma)$,
and hence serves as an upper bound as well.

\section*{Appendix A}
\renewcommand{\theequation}{A.\arabic{equation}}
    \setcounter{equation}{0}

In this appendix, we prove eq.\ (\ref{expectedempiricalentropy}). The proof is very similar to the proof
of eq.\ (32) in \cite{me14}, but here we derive a tighter redundancy term by exploiting the fact that
the number of distinct pairs $\{(u^\ell,x^\ell)\}$, that may appear as non--overlapping $\ell$-blocks of $(u^n,x^n)$,
cannot really be as large as $(\alpha\beta)^\ell$, 
but only $s_{\mbox{\tiny e}}\cdot\alpha^\ell$ at most. The simple reason is that $x^\ell$ is a 
deterministic function of $(z^{\mbox{\tiny e}},u^\ell)$, which in turn takes on at most 
$s_{\mbox{\tiny e}}\cdot\alpha^\ell$ different values. Although the proof is very similar to that of \cite{me14},
we provide it here in full detail, for the sake of completeness.

As explained in \cite{me14}, we invoke
the following result (see \cite{Atteson99}, \cite{CB90} and [19, Proposition
5.2] therein, as well as \cite[Appendix A]{MW04}):
Let $\hat{P}_n$ be the first order
empirical distribution associated with an $n$--sequence drawn from a
memoryless $m$--ary source $P$ with alphabet $\{1,\ldots,m\}$. Then,
\begin{equation}
	n\cdot\bE\{D(\hat{P}_n\|P)\}=\frac{(m-1)\log e}{2}+o(1),
\end{equation}
which is equivalent to
\begin{equation}
	\bE\left\{-\sum_{k=1}^m \hat{P}_n(k)\log \hat{P}_n(k)\right\}
= H-\frac{(m-1)\log e}{2n}-o\left(\frac{1}{n}\right),
\end{equation}
where $H$ is the entropy of $P$.
We now apply this result to the `source'
$P(y^\ell|u^\ell,z^{\mbox{\tiny e}})\equiv P(y^\ell|x^\ell)$ for every pair
$(u^\ell,z^{\mbox{\tiny e}})$ that appears more than $\epsilon n/\ell$ times as
$\ell$--blocks along the (deterministic) sequence pair
$(u^n,x^n)$.
\begin{eqnarray}
	&&\bE H(\hY^\ell|\hU^\ell,\hZ^{\mbox{\tiny e}})\nonumber\\
	&=&\bE\left\{\sum_{u^\ell,z^{\mbox{\tiny e}}}P_{\hU^\ell
\hZ^{\mbox{\tiny e}}}(u^\ell,z^{\mbox{\tiny e}})H(\hY^\ell|\hU^\ell=u^\ell,\hZ^{\mbox{\tiny e}}=z^{\mbox{\tiny e}})\right\}\nonumber\\
&\ge&\sum_{\{u^\ell,z^{\mbox{\tiny e}}:~P_{\hU^\ell\hZ^{\mbox{\tiny e}}}(u^\ell,z^{\mbox{\tiny e}})
	\ge\epsilon\}}P_{\hU^\ell
	\hZ^{\mbox{\tiny e}}}(u^\ell,z^{\mbox{\tiny e}})\cdot\bE\{H(\hY^\ell|\hU^\ell=
	u^\ell,\hZ^{\mbox{\tiny e}}=z^{\mbox{\tiny e}})\}\nonumber\\
&=&\sum_{\{u^\ell,z^{\mbox{\tiny e}}:~P_{\hU^\ell\hZ^{\mbox{\tiny e}}}(u^\ell,z^{\mbox{\tiny e}})\ge\epsilon\}}P_{\hU^\ell
	\hZ^{\mbox{\tiny e}}}(u^\ell,z^{\mbox{\tiny e}})\left[H(Y^\ell|\hX^\ell=q(z^{\mbox{\tiny e}},u^\ell))-
	\frac{(\gamma^\ell-1)\log
e}{2nP_{\hU^\ell
\hZ^{\mbox{\tiny e}}}(u^\ell,z^{\mbox{\tiny e}})/\ell}-o\left(\frac{\ell}{n\epsilon}\right)\right]\nonumber\\
&\ge&\sum_{\{u^\ell,z^{\mbox{\tiny e}}:~P_{\hU^\ell\hZ^{\mbox{\tiny e}}}(u^\ell,z^{\mbox{\tiny e}})
	\ge\epsilon\}}P_{\hU^\ell
\hZ^{\mbox{\tiny e}}}(u^\ell,z^{\mbox{\tiny e}})H(Y^\ell\|\hX^\ell=q(z^{\mbox{\tiny e}},u^\ell))-
	\frac{\ell s_{\mbox{\tiny e}}(\alpha\gamma)^\ell
\log e}{2n}
-o\left(\frac{\ell}{n\epsilon}\right)\nonumber\\
&=&\sum_{u^\ell,z^{\mbox{\tiny e}}}P_{\hU^\ell
\hZ^{\mbox{\tiny e}}}(u^\ell,z^{\mbox{\tiny e}})H(Y^\ell|\hX^\ell=q(z^{\mbox{\tiny e}},u^\ell))-\nonumber\\
	& &\sum_{\{u^\ell,x^\ell:~P_{\hU^\ell\hX^\ell}(u^\ell,x^\ell)<\epsilon\}}P_{\hU^\ell
	\hZ^{\mbox{\tiny e}}}(u^\ell,z^{\mbox{\tiny e}})H(Y^\ell|\hX^\ell=q(z^{\mbox{\tiny e}},u^\ell))
	-\frac{\ell s_{\mbox{\tiny e}} (\alpha\gamma)^\ell
\log e}{2n}-
o\left(\frac{\ell}{n\epsilon}\right)\nonumber\\
&\ge&
	H(Y^\ell|\hX^\ell)-\epsilon s_{\mbox{\tiny e}}\alpha^\ell\cdot \ell\log \gamma
	-\frac{\ell s_{\mbox{\tiny e}} (\alpha\gamma)^\ell
\log e}{2n}-
o\left(\frac{\ell}{n\epsilon}\right)\nonumber\\
	&\dfn&H(Y^\ell|\hX^\ell)-\ell\cdot \Delta_0(s_{\mbox{\tiny e}},\epsilon,\ell,n).
\end{eqnarray}
Selecting $\epsilon=1/\sqrt{n}$, we have
\begin{equation}
\Delta_1(s_{\mbox{\tiny e}},\ell,n)=\Delta_0\left(s_{\mbox{\tiny e}},\frac{1}{\sqrt{n}},\ell,n\right)=
	\frac{s_{\mbox{\tiny e}}\alpha^\ell\log\gamma}{\sqrt{n}}+
o\left(\frac{1}{\sqrt{n}}\right),
\end{equation}
as defined in eq.\ (\ref{delta1}). This completes the proof of eq.\ (\ref{expectedempiricalentropy}).

\section*{Appendix B}
\renewcommand{\theequation}{B.\arabic{equation}}
    \setcounter{equation}{0}

In this appendix, we prove that the function $\Psi(\cdot)$, defined in (\ref{psi}), is the inverse of the function
$\Phi(D)$, defined in (\ref{phi}). Let us denote
\begin{equation}
	R=\Phi(D)=\sup_{\theta\ge 0}\left[\theta D+\log\left(\sum_{u\in\calU}2^{-\theta\varrho(u)}\right)\right].
\end{equation}
This means that:
\begin{enumerate}
	\item $\forall~\theta\ge 0$, 
		\begin{equation}
	R\ge \theta D+\log\left(\sum_{u\in\calU}2^{-\theta\varrho(u)}\right).
		\end{equation}
	\item There exists a positive sequence, $\{\theta_n\}$, such that
		\begin{equation}
			\lim_{n\to\infty}\left[\theta_n D+\log\left(\sum_{u\in\calU}2^{-\theta_n\varrho(u)}\right)\right]=R.
		\end{equation}
\end{enumerate}
But this is clearly equivalent to the set of statements:
\begin{enumerate}
        \item $\forall~\theta\ge 0$,
                \begin{equation}
			D\le \frac{R-\log\left(\sum_{u\in\calU}2^{-\theta\varrho(u)}\right)}{\theta}.
                \end{equation}
        \item $\exists~\theta\ge 0$,
                \begin{equation}
			\lim_{n\to\infty}\frac{R-\log\left(\sum_{u\in\calU}2^{-\theta_n\varrho(u)}\right)}{\theta_n}=D.
                \end{equation}
\end{enumerate}
This in turn is equivalent to the statement that
\begin{equation}
	D=\inf_{\theta\ge 0}\frac{R-\log\left(\sum_{u\in\calU}2^{-\theta\varrho(u)}\right)}{\theta}
	=\inf_{\vartheta\ge 0}\vartheta\cdot\left[R-\log\left(\sum_{u\in\calU}2^{-\varrho(u)/\vartheta}\right)\right]=\Psi(R).
\end{equation}


\begin{thebibliography}{AA}

\bibitem{Atteson99}
K.~Atteson, ``The asymptotic redundancy of Bayes
rules for Markov chains,''
{\it IEEE Trans.\ Inform.\ Theory}, vol.\ 45,
no.\ 6, pp.\ 2104--2109, September 1999.

\bibitem{CB90}
B.~S.~Clarke and A.~R.~Barron, ``Information-theoretic asymptotics of 
Bayes methods,'' {\em IEEE Trans.\ Inform.\ Theory}, vol.\ 36, pp.\ 453-471,
May 1990.

\bibitem{CT06}
T.~M.~Cover and J.~A.~Thomas, {\it Elements of Information Theory}, John Wiley \& Sons, 
Hoboken, New Jersey, 2006.

\bibitem{Csiszar80}
I.~Csisz\'ar, ``Joint source--channel error exponent,'' {\em Problems of
Control and Information Theory}, vol.\ 9, no.\ 5, pp.\ 315--328, 1980.

\bibitem{Csiszar82}
I.~Csisz\'ar, ``On the error exponent of source-channel transmission with
a distortion threshold,''
{\em IEEE Trans.~Inform.~Theory\/},
vol.~IT--28, no.~6, pp.~823--828, November 1982.

\bibitem{EGK11}
A.~El Gamal and Y.-H.~Kim, {\it Network Information Theory}, Cambridge University Press, Cambridge, 2011.

\bibitem{Gallager68}
R.~G.~Gallager, {\em Information Theory and Reliable Communication}, John Wiley \& Sons, New York, 1968.

\bibitem{GP80}
S.~I.~Gel'fand and M.~S.~Pinsker, ``Coding for channel with random 
parameters,'' {\em Problems of Information and Control}, vol.\ 9, no.\ 1, pp.
19-31, 1980.

\bibitem{Gray73}
R.~M.~Gray, ``A new class of lower bounds to information rates of
stationary sources via conditional rate-distortion functions,'' {\em IEEE Trans.\
inform.\ Theory}, vol.~IT-19, no.~4, pp.~480--489, July 1973.

\bibitem{KSM07}
G.~Keshet, Y.~Steinberg, and N.~Merhav,
``Channel coding in the presence of side
information,'' {\it Foundations and Trends in
Communications and Information Theory}, vol.\ 4, no.\ 6,
pp.\ 445--586, 2007.

\bibitem{LMW14}
A.~Lapidoth, A.~Mal\"ar, and M.~Wigger, ``Constrained source--coding with side
information,'' {\em IEEE Trans.\ Inform.\ Theory\/},
vol.\ 66, no.~6, pp.~3218--3237, June 2014.

\bibitem{LZ76}
A.~Lempel and J.~Ziv, ``On the complexity of finite sequences,'' 
{\em IEEE Trans.\ Inform.\ Theory\/},
vol.~IT--22, no.~1, pp.~75--81, January 1976.

\bibitem{LZ86}
A.~Lempel and J.~Ziv, ``Compression of two--dimensional data,''
{\em IEEE Trans.\ Inform.\ Theory}, vol.\
IT--32, no.\ 1, pp.\ 2--8, January 1986.

\bibitem{Marton74}
K.~Marton, ``Error exponent for source coding with a fidelity 
criterion,'' {\em IEEE Trans.~Inform.~Theory\/},
vol.~IT--20, no.~2, pp.~197--199, March 1974.

\bibitem{me00}
N.~Merhav, ``Universal detection of messages via finite--state channels,''
{\em IEEE Trans.\ Inform.\ Theory},
vol.\ 46, no.\ 6, pp.\ 2242--2246, September 2000.

\bibitem{me14}
N.~Merhav, ``On the data processing theorem in the semi--deterministic
setting,'' {\em IEEE Trans.\ Inform.\ Theory},
vol.\ 60, no.\ 10, pp.\ 6032--6040, October 2014.

\bibitem{MS03}
N.~Merhav and S.~Shamai (Shitz),
``On joint source--channel coding for the Wyner--Ziv source and
the Gel'fand--Pinsker channel,''
{\em IEEE Trans.\ Inform.\ Theory},
vol.\ 49, no.\ 11, pp.\ 2844--2855, November 2003.

\bibitem{MW04}
N.~Merhav and M.~J.~Weinberger, ``On universal simulation of information
sources using training data,''
{\it IEEE Trans.\ Inform.\ Theory},
vol.\ 50, no.\ 1, pp.\ 5--20, January 2004.

\bibitem{MZ06}
N.~Merhav and J.~Ziv, ``On the Wyner--Ziv problem for individual sequences,''
{\em IEEE Trans.\ Inform.\ Theory}, vol.\ 52, no.\ 3, pp.\ 867--873, March
2006.

\bibitem{SVZ98}
S.~Shamai (Shitz), S.~Verd\'u and R.~Zamir, ``Systematic lossy source/ 
channel coding,'' {\em IEEE Trans.~Inform.~Theory\/},
vol.\ 44, no.\ 2, pp.\ 564--579, March 1998.

\bibitem{Shannon58}
C.~E.~Shannon, ``Channels with side information at the transmitter,'' {\em IBM
Journal Research and Development}, vol.\ 2, pp.\ 289-293, October 1958.

\bibitem{SGB67}
C.~E.~Shannon, R.~G.~Gallager, and E.~R.~Berlekamp, ``Lower bounds to error probability for coding in discrete memoryless channels,''
{\em Inform.\ Contr.}, vol.\ 10, pp.\ 65--103 and 522--552, 1967.

\bibitem{Steinberg09}
Y.~Steinberg, ``Coding and common reconstruction,'' {\em IEEE Trans.\ Inform.\ Theory}, vol.~55,
no.~11, pp.~4995-–5010, November 2009.

\bibitem{WZ76}
A.~D.~Wyner and J.~Ziv, ``The rate-distortion function for source 
coding with side information at the decoder,''
{\em IEEE Trans.~Inform.~Theory\/},
vol.~IT--22, no.~1, pp.~1--10, January 1976.

\bibitem{Zamir96}
R.~Zamir, ``The rate loss in the Wyner--Ziv problem,''  {\em IEEE Trans.\ Inform.\ Theory}, vol.~42, no.\ 6,
pp.\ 2073--2084, November 1996.

\bibitem{Ziv78}
J.~Ziv, ``Coding theorems for individual sequences,'' 
{\em IEEE Trans.~Inform.~Theory\/},
vol.~IT--24, no.~4, pp.~405--412, July 1978.

\bibitem{Ziv80}
J.~Ziv, ``Distortion--rate theory for individual sequences,'' 
{\em IEEE Trans.~Inform.~Theory\/},
vol.~IT--26, no.~2, pp.~137--143, March 1980.

\bibitem{Ziv84}
J.~Ziv, ``Fixed-rate encoding of individual sequences with side 
information,'' {\em IEEE Trans.\ Inform.\ Theory\/},
vol.~IT--30, no.~2, pp.~348--452, March  1984.

\bibitem{Ziv85}
J.~Ziv, ``Universal decoding for finite-state channels,'' 
{\em IEEE Trans.~Inform.~Theory\/},
vol.~IT--31, no.~4, pp.~453--460, July 1985.

\bibitem{ZL78}
J.~Ziv and A.~Lempel, ``Compression of individual sequences via 
variable-rate coding,''
{\em IEEE Trans.~Inform.~Theory\/},
vol.~IT--24, no.~5, pp.~530--536, September 1978.

\end{thebibliography}
\end{document}